# Implementation of Handoff through wireless access point Techniques

N.S.V.Shet, K.Chandrasekaran and K.C.Shet

**Abstract**— Handoff has become an inevitable part of wireless cellular communication, Soon users will carry small portable handheld devices which will incorporate the computer, phone, camera, GPS, personal control module etc. This paper proposes a new scheme to deal with seam less roaming and reduce failed handoffs. The simulation is done using software called Qualnet meant for wireless communication. The results clearly indicate the advantages of this new scheme.

——————————— ◆ ———————————

## 1 INTRODUCTION

In the present-day scenario, it is observed that several handoff events take place, while mobile user's move short distances, thereby causing poor QoS and undue load on base station processors. Service providers would welcome the possibility of reducing the no. of base stations during deployment of mobile services. The outcome of reduced handoff would result in ultimately improving the QoS and there by suggesting reduction in no. of base stations. It is also observed that when a mobile device such as a laptop is moved from one access point to another it needs to reconfigure itself, usually by human intervention. In situations where the event is VOIP or video streaming, then such a hindrance is not welcome.

The continuation of an active call is one of the most important quality measurements in cellular systems. Handoff process enables a cellular system to provide such a facility by transferring an active call from one cell to another. Different approaches are proposed and applied in order to achieve better handoff service. The principal parameters used to evaluate handoff techniques are, forced termination and call blocking probability. Mechanisms such as guard channels and queuing handoff calls decrease the forced termination probability while increasing the call blocking probability. A good amount of research is done, in suggesting handoff algorithms. However, apparently, few have tried to simulate a handoff scenario and obtain results to improve QoS.

A simulation-based study for handoff is required and it is anticipated to improve QoS. Future campuses such as industry or academic will be equipped with access points at every corner to facilitate wireless Internet connectivity. These wireless access points can act as nano base stations such that they can route calls/messages to the service provider's base stations. Several routing algorithms/ hand shake algorithms are already in place and these can be used for such access. It is also evident that most of the mobile phone users remain indoors during most of their working hours and hence connectivity through access points seems quite reasonable.

## 2 ABOUT WIRELESS ACCESS POINTS

In computer networking, a wireless access point (WAP or AP) is a device that allows wireless communication devices to connect to a wireless network using Wi-Fi , Bluetooth or related standards. The WAP usually connects to a wired network, and can relay data between the wireless devices (such as computers or printers) and wired devices on the network. Prior to wireless networks, setting up a computer network in a business, home, or school often required running many cables through walls and ceilings in order to deliver network access to all of the network-enabled devices in the building. With the advent of the Wireless Access Point, network users are now able to add devices that access the network with few or NO cables. Today's WAP's are built to support a standard for sending and receiving data using radio frequencies rather than cabling. Those standards and the frequencies they use are defined by the IEEE. Most WAP's use IEEE 802.11 standards.

- *N.S.V.Shet is with NITK,* Surathkal, India
- *K.Chandrasekaran is with NITK,* Surathkal, India
- *K.C.Shet is with NITK,* Surathkal, India

A typical corporate use involves attaching several WAP's to a wired network and then providing wireless access to the office LAN. Within the range of the WAP's, the wireless end user has a full network connection with the benefit of mobility. In this instance, the WAP functions as a gateway for clients to access the wired network. A WAP may also act as the network's arbitrator, negotiating when each nearby client device can transmit. However, the vast majority of currently installed IEEE 802.11 networks do not implement this, using a distributed pseudo-random algorithm called CSMA/CA instead.

## 3 HANDOFF SCHEME WITH WIRELESS ACCESS POINTS

The scheme uses wireless access points (WAP) as the ad-hoc routing station, to connect to the base stations. Wherever a base station access for handoff is not feasible the call is handed over to the wireless access point and call continuity is maintained.

The flow in appendix searches for signals of wireless access points in the vicinity. If the no of access points are more, then it chooses the best signal access point. Once the access point is decided it looks at the type of handoff, like soft or hard handoff and accordingly hands over the call. Eventually it also has certain threshold settings for no of attempts before handoff decision etc.

The scene is created in Qualnet and consists of two base stations, few WAP's, one mobile stations and one Mobile switching centre. The scene is designed such that the mobile stations which are initially near one of the base stations moves in a circular path to finally come back to the original BS. As the simulation progresses it moves such that it passes near the wireless access points to finally reach the opposite base station. Signal exchanges take place which is clear from the simulations. The result of packet exchanges in the various OSI layers is also shown.

Scenarios to demonstrate WAP handoff technique

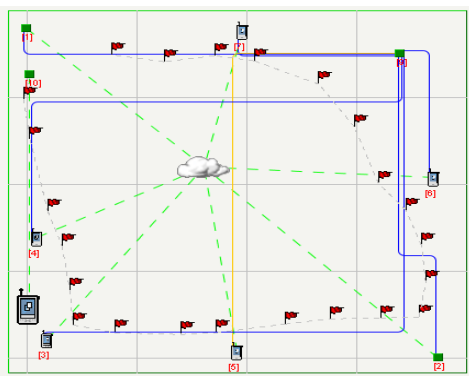

About the scene created

In fig Node 1 & 2 represent Base stations, node 9 is a mobile switching center, node 3, 4, 5, 6 and 7 behave as access points, and node 10 is a mobile station. The mobile station broadcasts the packets as per 802.11 MAC std. and the base stations receive them. The triggered updates sent as per the Bellman-ford algorithm is highlighted. The mobile station moves in a circular path such that the signals from base station are lost while the signals from the access points can be accessed.

Results of WAP handoff technique simulation

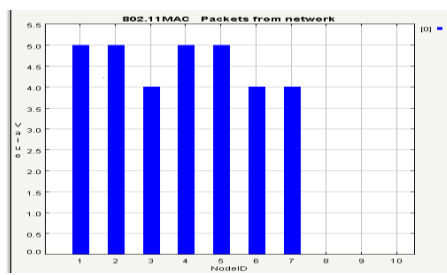

Y axis – signal strength depends upon technology used

Y axis - no of packets

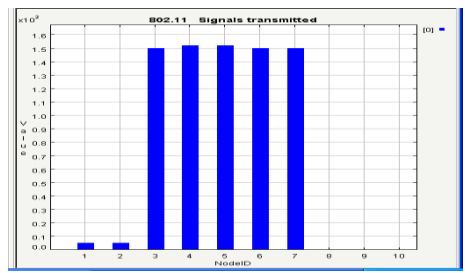

Y axis – ip In addr received

The results show various packet transfers in the different OSI layers during simulation.

The changes in various parameters in the QSI layers with wireless access points and without wireless access

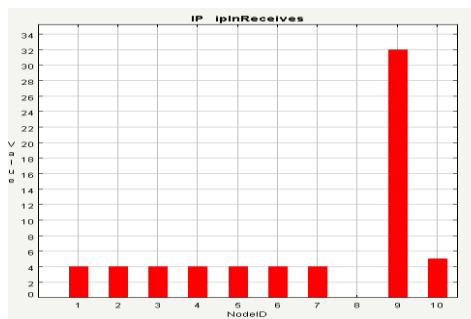

points are listed below.

Physical Layer (802.11)

Signal strength required is 1/100, for the case with access point in comparison with case without access point . (0.05 units and 5 units)
Signal Rx & Fwd to MAC is reduced from 3 units to 1.02 units
Signal locked on by Physical layer reduced from 4 units to 1.73 units
Signal Rx but with errors reduced from 1 unit to 0.73 units.

MAC Layer (802.11 MAC)

No. of broadcast packets received clearly increased from 3 to 5.

MAC Layer 802.11 DCF (Distributed co-ordinate function)

No. of broadcast packets received increased from 3 to 5.

MAC Link Layer

Number of Link Frames sent is constant at 4 for each access point.
Number of Link frames received is less than or equal to 5.
Link utilization almost doubled from 0.000012 to 0.000025.

**Network Layer IP**
Mobile Station ip In Rx improved from 10 units to 32 units , ip In delivers from 10 units to to 32, ip Out requests improved from 8 units to 28 units.

MS ip In delivers TTL Sum reduced from 1.8 units to 0.3 units and for MSC from 6.25 units to 2.02 units – **undesirable.**

**Network Layer Strict Prior**

**No. of packet queuing at MSC increased from 8 to 28 - undesirable**

**Network Layer FIFO**
Total packets queued and dequeued increased from 8 to 28 -**undesirable**
Packet size increased **– undesirable.**
Peak queue size is nil.

**Transport Layer UDP**
No. of packets from and to application layer increased from 8 to 28 and 10 to 32. respectively - **undesirable**

**Application Layer Bellman Ford**
No. of update packets received increased from 10 to 32.

## Conclusion

Among the various parameters analyzed and compared with and without wireless access Points, 8 parameters have exhibited improvement in performance, 4 parameters have exhibited undesirable change, 5 parameters have exhibited change which is not significant in performance evaluation. Hence an improvement by 66.66 % of Q.o.S is observed. Using WAP techniques to provide handoff as well as connectivity in urban environment where signals are difficult to reach is a feasible alternative. Installation of expensive base stations can be avoided. In the future access points will be placed in all most every corner of an office building therefore uninterrupted handoff and hence seamless mobility can be achieved in an urban environment.

## REFERENCES

1.Wikipideaencyclopedia@http://en.wikipedia.org/wiki/Wireless_access_point

2. N.S.V.Shet, *"Newer algorithms for handoff in wireless mobile communication using wireless sensor node techniques"* Proceedings of KU-NITK joint seminar, 27th-28th Nov 2008, Kagoshima University, Japan, pp 66-72.

3. S. Limm, G. Cao, C. R. Das, *"A differential bandwidth reservation policy for multimedia wireless networks"*, Parallel Processing Workshops, International Conference on, 2001, pp. 447-452.

4. D.A. Levine, I.F. Akyildiz, M. Naghshineh "A resource, estimation and call admission algorithm for wireless multimedia networks using the shadow cluster concept". IEEE/ACM Transactions Networking **5** (1997), pp. 1- 12.

5. Cross bow technologies ,USA Mote view User Manualat
http://www.xbow.com/Products/Product_pdf_files/Wireless_pdf/MoteWorks_OEM_Edition.pdf

6. Product tour and user's manual of Qualnet 4.5 by Scalable technologies, USA.

## APPENDIX

Scheme Flow diagram

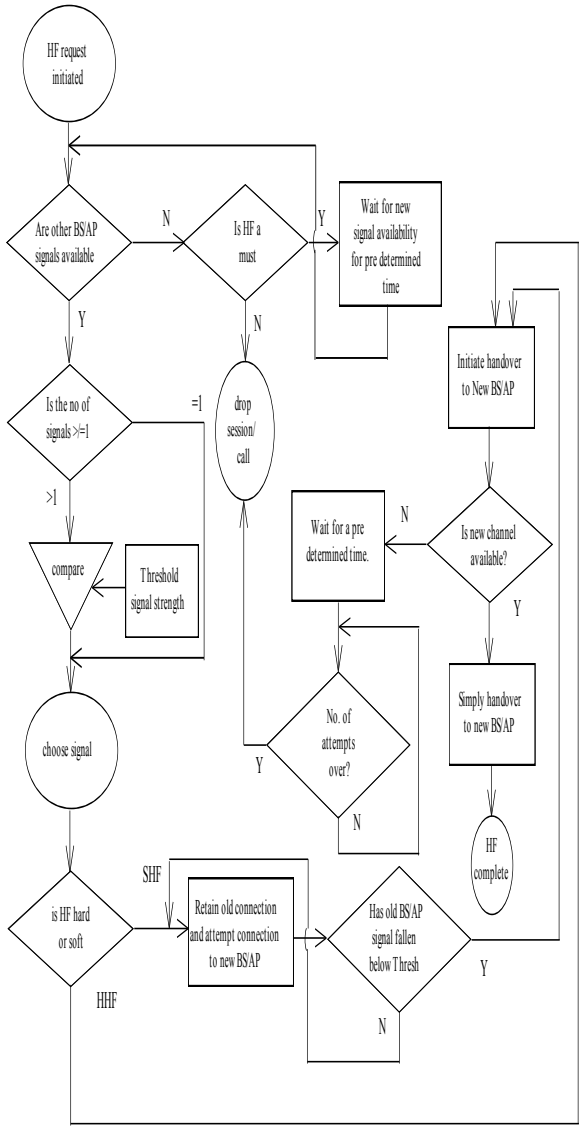